\documentclass[prl,aps,preprint,showpacs]{revtex4}
\usepackage{epsfig,amssymb,amsfonts}

\def\opone{\leavevmode\hbox{\small1\kern-3.8pt\normalsize1}}

\begin{document}

\title{Thermal entanglement in a two-qubit Heisenberg \emph{XXZ} spin chain under an inhomogeneous magnetic field}

\author{Guo-Feng Zhang\footnote{Corresponding author; electronic address:
gf1978zhang2001@yahoo.com}} \affiliation{State Key Laboratory for
Superlattices and Microstructures, Institute of Semiconductors,
Chinese Academy of Sciences, P. O. Box 912, Beijing 100083, P. R.
China}
\author{Shu-Shen Li}
\affiliation{China Center of Advanced Science and Technology
(CCAST) (World Laboratory), P.O. Box 8730, Beijing 100080, China
and State Key Laboratory for Superlattices  and Microstructures,
Institute of Semiconductors, Chinese Academy of Sciences, P.O. Box
912, Beijing 100083,China}

\begin{abstract}
The thermal entanglement in a two-qubit Heisenberg \emph{XXZ} spin
chain is investigated under an inhomogeneous magnetic field
\emph{b}.
 We show that the ground-state entanglement is independent
of the interaction of \emph{z}-component $J_{z}$. The thermal
entanglement at the fixed temperature can be enhanced when $J_{z}$
increases. We strictly show that for any temperature \emph{T} and
$J_{z}$ the entanglement is symmetric with respect to zero
inhomogeneous magnetic field, and the critical inhomogeneous
magnetic field $b_{c}$ is independent of $J_{z}$. The critical
magnetic field $B_{c}$ increases with the increasing $|b|$ but the
maximum entanglement value that the system can arrive becomes
smaller.
\end{abstract}

\pacs{03.67.Lx; 03.65.Ud; 75.10.Jm; 05.50.+q } \maketitle
\maketitle

\section{I. Introduction} Entanglement is the most fascinating features of quantum
mechanics and plays a central role in quantum information
processing. In recent years, there has been an ongoing effort to
characterize qualitatively and quantitatively the entanglement
properties of condensed matter systems and apply then in quantum
communication and information. An important emerging field is the
quantum entanglement in solid state systems such as spin chains
\cite{man,xwa,glk,kmo,ysu,yye,dvk,gfz}. Spin chains are the
natural candidates for the realization of the entanglement
compared with the other physics systems. The spin chains not only
have useful applications such as the quantum state transfer, but
also display rich entanglement features \cite{sbo}. The Heisenberg
chain, the simplest spin chain, has been used to construct a
quantum computer and quantum dots \cite{dlo}. By suitable coding,
the Heisenberg interaction alone can be used for quantum
computation \cite{dal,dpd,lfs}. The thermal entanglement, which
differs from the other kinds of entanglements by its advantages of
stability for the reduction in entanglement of an entangled state
due to various sources of decoherence and in entanglement in time
due to thermal interactions are absent as the entanglement at
finite temperature takes thermal decoherence into account
implicitly, requires neither measurement nor controlled switching
of interactions in the preparing process, and hence becomes an
important quantity of systems for the purpose of quantum
computing. In the studies on the entanglement of Heisenberg spin
model, a lot of interesting work have been done
\cite{xgw,shi,mca}. It is turned out that the critical magnetic
field $B_{c}$ is increased by introducing the interaction of the
$z$-component of two neighboring spin in Ref [17]. But only the
uniform field case is carefully studied in the above-mentioned
papers. The nonuniform case is rarely taken into account. We know
that in any solid state construction of qubits, there is always
the possibility of inhomogeneous zeeman coupling \cite{xhua,xhul}.
So it is necessary to consider the entanglement for a nonuniform
field case. M.Asoudeh and V.Karimipour \cite{mas} studied the
effect of inhomogeneous in the magnetic field on the thermal
entanglement of an isotropic two-qubit \emph{XXX} spin system. We
noticed that the entanglement for a \emph{XXZ} spin model in an
nonuniform field has not been discussed. Although M.Asoudeh
\cite{mas} states that the different types of anisotropic
interactions may not be of much practical relevance to concrete
physical realization of qubits, in the theoretical analysis we
think it is very interesting and should be included in the studies
of spin chain entanglement. This is the main motivation of this
paper.

   For a system in equilibrium at temperature $T$, the density
matrix is $\rho=\frac{1}{Z}e^{-\beta\emph{H}}$, where
$\beta=\frac{1}{k\emph{T}}$ , $k$ is the Boltzman constant and
$Z=tre^{-\beta\emph{H}}$ is the partition function. For
simplicity, we write $k=1$. The entanglement of two qubits can be
measured by the concurrence $C$ which is written as $C=\max[0,2
\max[\lambda_{i}]-\sum^{4}_{i}\lambda_{i}]$\cite{shi}, where
$\lambda_{i}$ are the square roots of the eigenvalues of the
matrix $R=\rho S\rho^{*}S$, $\rho$ is the density matrix,
$S=\sigma_{1}^{y}\otimes\sigma_{2}^{y}$ and $*$ stands for the
complex conjugate. The concurrence is available, no matter whether
$\rho$ is pure or mixed. In case that the state is pure
$\rho=|\psi\rangle\langle\psi|$ with
\begin{equation}
|\psi\rangle=a|0,0\rangle+b|0,1\rangle+c|1,0\rangle+d|1,1\rangle,
\end{equation}
the concurrence is simplified to
\begin{equation}
C(\psi)=2|ad-bc|.
\end{equation}

\section{II. The model and the Ground-state entanglement}
The Hamiltonian of the \emph{N}-qubit anisotropic
Heisenberg \emph{XXZ} model under an inhomogeneous magnetic field
is
\begin{eqnarray}
\emph{H}&=&\frac{1}{2}\sum^{N}_{i=1}[J\sigma_{i}^{x}\sigma_{i+1}^{x}+
J\sigma_{i}^{y}\sigma_{i+1}^{y}
+J_{z}\sigma_{i}^{z}\sigma_{i+1}^{z}
+(B+b)\sigma_{i}^{z}+(B-b)\sigma_{i+1}^{z}],
\end{eqnarray}
where $J$ and $J_{z}$ are the real coupling coefficients. The
coupling constant $J>0$ and $J_{z}>0$ corresponds to the
antiferromagnetic case, $J<0$ and $J_{z}<0$ the ferromagnetic
case. $B\geq0$ is restricted, and the magnetic fields on the two
spins have been so parameterized that \emph{b} controls the degree
of inhomogeneity. Now, we consider the Hamiltonian for $N=2$ case.
Note that we are working in units so that $B$, $b$, $J$ and
$J_{z}$ are dimensionless.

In the standard basis
$\{|1,1\rangle,|1,0\rangle,|0,1\rangle,|0,0\rangle\}$, the
Hamiltonian can be expressed as
\begin{eqnarray}
\emph{H}=
\left(%
\begin{array}{cccc}
  \frac{J_{z}+2B}{2} & 0 &0 &0 \\
 0 & \frac{-J_{z}+2b}{2} & J & 0\\
  0 & J & \frac{-J_{z}-2b}{2} &0 \\
 0 & 0 & 0 & \frac{J_{z}-2B}{2} \\
\end{array}%
\right),
\end{eqnarray}

A straightforward calculation gives the following eigenstates:
\begin{eqnarray}
&|&\phi_{1}\rangle=|0,0\rangle,\nonumber
\\&|&\phi_{2}\rangle=|1,1\rangle,\nonumber
\\&|&\phi_{3}\rangle=\frac{1}{\sqrt{1+\xi^{2}/J^{2}}}(\frac{\xi}{J}|1,0\rangle+|0,1\rangle),\nonumber
\\&|&\phi_{4}\rangle=\frac{1}{\sqrt{1+\zeta^{2}/J^{2}}}(\frac{\zeta}{J}|1,0\rangle+|0,1\rangle),
\end{eqnarray}
with corresponding energies:
\begin{eqnarray}
&E_{1}&=\frac{1}{2}(J_{z}-2B),\nonumber
\\&E_{2}&=\frac{1}{2}(J_{z}+2B),\nonumber
\\&E_{3}&=-\frac{J_{z}}{2}-\eta,\nonumber
\\&E_{4}&=-\frac{J_{z}}{2}+\eta.
\end{eqnarray}
where $\eta=\sqrt{b^{2}+J^{2}}$, $\xi=b-\eta$ and $\zeta=b+\eta$.
Note that when $b\rightarrow0$ and $J>0$ , the two states
$|\phi_{3}\rangle$ and $|\phi_{4}\rangle$ respectively go to the
maximally entangled state
$\frac{1}{\sqrt{2}}(|0,1\rangle-|1,0\rangle)$ and
$\frac{1}{\sqrt{2}}(|0,1\rangle+|1,0\rangle)$. For $J<0$ , they
respectively go to $\frac{1}{\sqrt{2}}(|0,1\rangle+|1,0\rangle)$
and $\frac{1}{\sqrt{2}}(|0,1\rangle-|1,0\rangle)$. We can also
find that the eigenenergies are even function of the coupling
constant $J$. So we can think the ground-state entanglement exists
for both antiferromagnetic and ferromagnetic cases and should be
symmetric with respect to the coupling constant $J$. The ground
state depends on the value of the magnetic field \emph{B} , the
coupling constant $J_{z}$ and $\eta$. It is readily found that the
ground-state energy is equal to
\begin{eqnarray}
\left\{%
\begin{array}{ll}
     E_{1}=\frac{1}{2}(J_{z}-2B), & \hbox{if $\eta<B-J_{z}$;} \\
    E_{3}=-\frac{J_{z}}{2}-\eta, & \hbox{if $\eta>B-J_{z}$.} \\
\end{array}%
\right.
\end{eqnarray}
So when $\eta<B-J_{z}$, the ground state is the distangled state
$|\phi_{1}\rangle$ and when $\eta>B-J_{z}$, the ground state is
the entangled state $|\phi_{3}\rangle$. For each value of the
magnetic field \emph{B} , there is a threshold parameter
$J_{z}^{f}=B-\eta$ above which the ground state will become
entangled. Accordingly for each value of inhomogeneity $\eta$
there is a value of magnetic field $\emph{B}^{f}=\eta+J_{z}$ above
which the ground state will loose its entanglement. In the
entangled phase the entanglement of the ground state is found from
(2) and (5) to be
\begin{equation}
C(|\phi_{3}\rangle)=\frac{2|\lambda|}{1+\lambda^{2}} ,
\end{equation}
where $\lambda=\xi/J$. $\lambda=\pm1$, i.e. $b=0$, the system
enters the maximally entangled phase $|\phi_{3}\rangle$ with
entanglement $C(|\phi_{3}\rangle)=1$. This result accords with
that in Ref\cite{mas}. Here we can also know that the ground-state
entanglement is independent of the interaction of
\emph{z}-component $J_{z}$.

\section{III. The thermal entanglement }As the thermal fluctuation is introducing into the system, the
entangled ground states will be mixed with the untangled excited
state. This effect will make the entanglement decreases. At the
same time, the distangled ground state mixes with entangled
excited states. To see the change of the entanglement, we
calculate the entanglement of the thermal state
$\rho=\frac{1}{Z}e^{-\beta\emph{H}}$. In the standard basis
$\{|1,1\rangle, |1,0\rangle, |0,1\rangle, |0,0\rangle\}$, the
density matrix of the system can be written as
\begin{eqnarray}
\rho_{12}=\frac{1}{Z}
\left(%
\begin{array}{cccc}
  e^{-\frac{E_{2}}{k T}}& 0  &0 & 0 \\
  0 & e^{\frac{J_{z}}{2kT}}(m-n) & -s & 0 \\
  0 &  -s & e^{\frac{J_{z}}{2kT}}(m+n) & 0\\
 0 &0 & 0 & e^{-\frac{E_{1}}{kT}} \\
\end{array}%
\right)
\end{eqnarray}
where $Z=e^{-\frac{E_{2}}{k
T}}(1+e^{\frac{2B}{kT}})+2e^{\frac{J_{z}+B}{kT}}\cosh\frac{\eta}{kT}$,
$m=\cosh\frac{\eta}{kT}$,
 $n=\frac{b\sinh\frac{\eta}{kT}}{\eta}$, $s=\frac{e^{\frac{J_{z}}{2kT}}J\sinh\frac{\eta}{kT}}{\eta}$.
In the following calculation, we will write the Boltzman constant
$k=1$. From Eq.(9) and the definition of concurrence, we can
obtain the concurrence at the finite temperature.

\emph{Case1: $J_{z}=0$.} Our model corresponds to a $XX$ spin
model. The eigenvalues and eigenvectors can be easily obtained. In
Fig.1, we give the results at different temperature for the
nonuniform magnetic field ($B=0$) and the uniform magnetic ($b=0$)
. From the figure, we can know that the entanglement is symmetric
with respect to zero magnetic field, the nonuniform magnetic field
can lead to higher entanglement and double-peak structure. This
results accord with those seen from Ref\cite{ysu}.
\begin{figure}
\begin{center}
\epsfig{figure=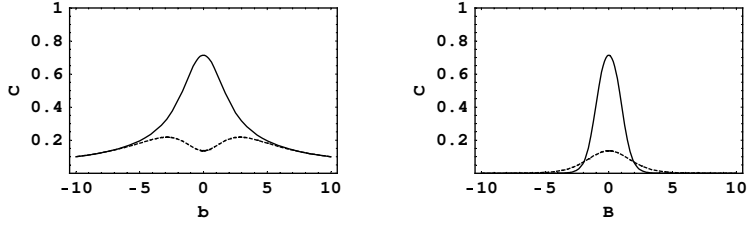,width=0.60\textwidth}
\end{center}
\caption{The concurrence for $J_{z}=0$ and $J=1$ case. $T=0.4$
(solid curve) and $T=1.0$ (dotted curve). The left panel
corresponds to the nonuniform case and the right panel corresponds
to the uniform case. $T$ is plotted in units of the Boltzmann's
constant $k$. And we work in units where $B$ and $b$ are
dimensionless.} \label{fig:FIG1}
\end{figure}

\emph{Case2: $J_{z}=J$.} In order to compare with the results in
Ref\cite{mas}, we make the substitution $J\rightarrow2J$ ,
$B\rightarrow2B$, $b\rightarrow2b$. The eigenvalues and
eigenvectors can be easily obtained as
\begin{eqnarray}
&|&\varphi_{1}\rangle=|0,0\rangle,\nonumber
\\&|&\varphi_{2}\rangle=|1,1\rangle,\nonumber
\\&|&\varphi_{3}\rangle=\frac{1}{\sqrt{1+x^{2}/J^{2}}}(\frac{x}{J}|1,0\rangle+|0,1\rangle),\nonumber
\\&|&\varphi_{4}\rangle=\frac{1}{\sqrt{1+y^{2}/J^{2}}}(\frac{y}{J}|1,0\rangle+|0,1\rangle),
\end{eqnarray}
with corresponding energies:
\begin{eqnarray}
&e_{1}&=J-2B,\nonumber
\\&e_{2}&=J-2B,\nonumber
\\&e_{3}&=-J-2|J|\sqrt{1+\delta^{2}},\nonumber
\\&e_{4}&=-J+2|J|\sqrt{1+\delta^{2}}.
\end{eqnarray}
where $\delta=b/J$, $x=2b-2\sqrt{1+\delta^{2}}$ and
$y=2b+\sqrt{1+\delta^{2}}$. For the ferromagnetic case $J=-1$, the
ground-state concurrence is
$C(|\varphi_{3}\rangle)=\frac{1}{\sqrt{1+\delta^{2}}}$. For the
ferromagnetic case $J=1$, the ground-state concurrence is
$C(|\varphi_{4}\rangle)=\frac{1}{\sqrt{1+\delta^{2}}}$. These
results are same with those obtained in Ref\cite {mas}. In Fig.2,
we give the plot of the thermal concurrence for $J_{z}=J=-1$ case.
\begin{figure}
\begin{center}
\epsfig{figure=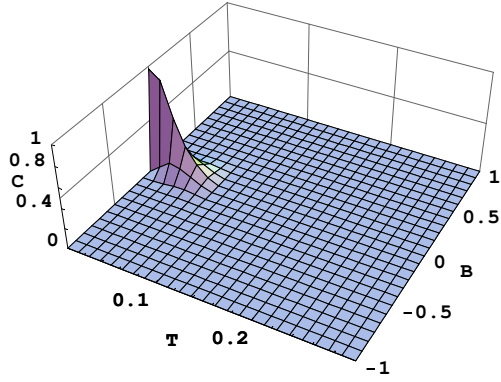,width=0.40\textwidth}
\end{center}
\caption{(Color online) The thermal concurrence for $J_{z}=J=-1$
case. And the inhomogeneous magnetic field $b=0.458$. $T$ is
plotted in units of the Boltzmann's constant $k$. And we work in
units where $B$ and $b$ are dimensionless.} \label{fig:FIG2}
\end{figure}
In order to compare our result with that in Ref\cite {mas}, we let
the inhomogeneous magnetic field $b=0.458$ (this accords to the
value of $\xi$ in Ref\cite {mas}). We can see that the thermal
entanglement develops and is maximized for zero magnetic field $B$
and gets the maximum value at $T=0$.

\emph{Case3: For any $J_{z}$.} With $B=0$, the concurrence as a
function of $b$ and $T$ for two value of $J_{z}$ are given in
Fig.3. They show that the concurrence are $1$ for different
$J_{z}$ when $b=0$ and $T=0$. At the point, the ground state is
$|\phi_{3}\rangle$ with energy $-\frac{J_{z}}{2}-1$, which is the
maximally entangled state and the corresponding concurrences are
1. As the temperature increases, the concurrences decreases due to
the mixing of other states with the maximally entangled state. We
can also know that the concurrence decrease with the increasing of
$|b|$. From the two figures in Fig.3, we can find that with the
increasing $J_{z}$, the critical temperature $T_{c}$ is improved
(for $J_{z}=0$, $T_{c}$ is about 2, but for $J_{z}=0.9$, $T_{c}$
has a higher value). Thus, we can get the higher entanglement at a
fixed temperature when $J_{z}$ is increased.
\begin{figure}
\begin{center}
\epsfig{figure=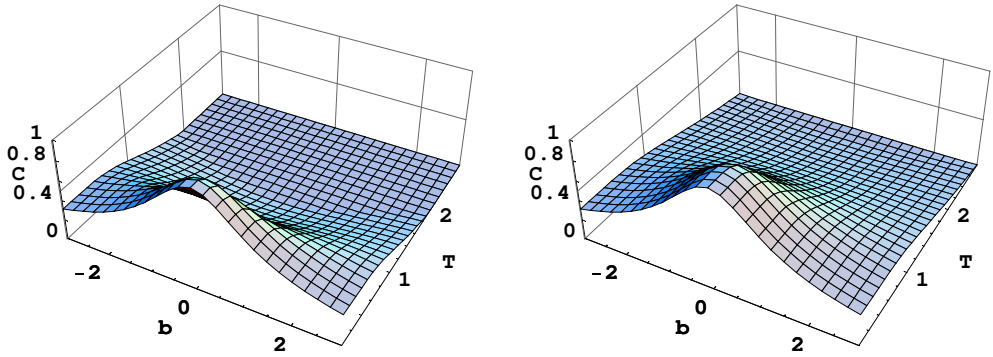,width=0.80\textwidth}
\end{center}
\caption{(Color online) The concurrence in the \emph{XXZ} spin
model is plotted vs \emph{b} and \emph{T}. Coupling constant
$J=1$, the magnetic field $B=0$. The left panel corresponds to
$J_{z}=0$ case and the right panel corresponds to $J_{z}=0.9$
case. $T$ is plotted in units of the Boltzmann's constant $k$. And
we work in units where $b$ is dimensionless. } \label{fig:FIG3}
\end{figure}

\begin{figure}
\begin{center}
\epsfig{figure=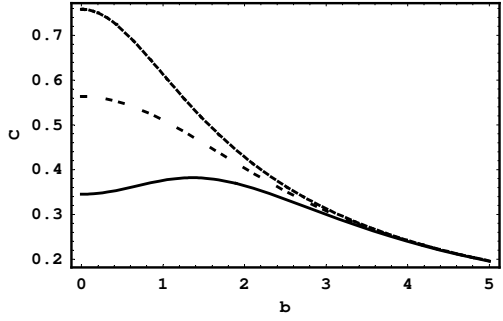,width=0.40\textwidth}
\end{center}
\caption{The concurrence in the \emph {XXZ} model is plotted vs
$b$ for various value of $J_{z}$, where $J=1$, $B=0.8$ and $T=0.6$
. From top to bottom, $J_{z}$ equals 0.9, 0.4, 0. $T$ is plotted
in units of the Boltzmann's constant $k$. And we work in units
where $B$ and $b$ are dimensionless.} \label{fig:FIG4}
\end{figure}
\begin{figure}
\begin{center}
\epsfig{figure=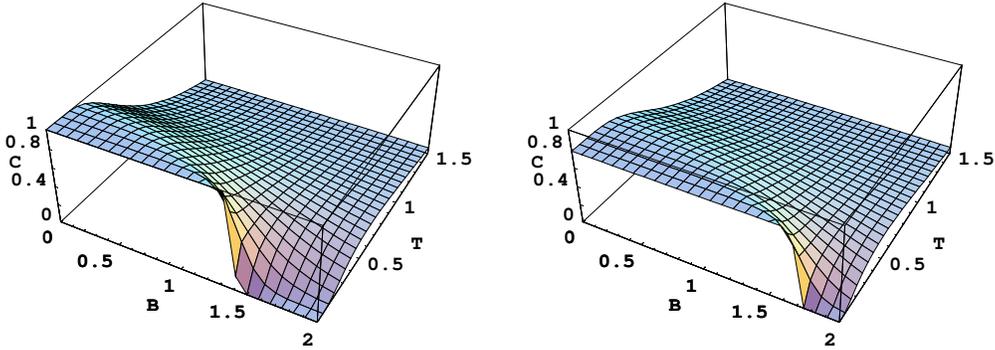,width=0.80\textwidth}
\end{center}
\caption{(Color online)The concurrence in the \emph{XXZ} spin
model is plotted vs $T$ and $B$, where $J=1$, $J_{z}=0.4$. The
left panel corresponds to $b=0$ case and the right panel
corresponds to $b=0.8$ case. $T$ is plotted in units of the
Boltzmann's constant $k$. And we work in units where $B$ and $b$
are dimensionless.} \label{fig:FIG5}
\end{figure}

Fig.4 shows the concurrence at a fixed temperature and magnetic
field for three values of positive $J_{z}$. It is shown that the
concurrences drop with the increasing value of $b$ and arrive to
zero at the same $b$ value, which is called the critical
inhomogeneous magnetic field, for various values of $J_{z}$. This
is to say the critical inhomogeneous magnetic field is independent
of $J_{z}$. Moreover, we can see that for a higher value of
$J_{z}$, the system has a stronger entanglement which is
consistent with Fig.3.

In Fig.5 we give the plot of concurrence as a function of $T$ and
$B$ for $b=0$ and $b=0.8$ respectively when $J_{z}=0.4$. For $B=0$
and $b=0$, the maximally entangled state
$|\Phi\rangle=\frac{1}{\sqrt{2}}(|0,1\rangle-|1,0\rangle)$ is the
ground state with eigenvalue $-\frac{J_{z}}{2}-|J|$. Then the
maximum entanglement is at $T=0$, i. e. $C=1$. As $T$ increases,
the concurrence decreases as seen from Fig.5 due to the mixing of
other states with the maximally entangled state. For a high value
of $B$ (in left Fig.5 $B=1.40$ and in right Fig.5 $B=1.70$) the
state $|\phi_{1}\rangle$ becomes the ground state, which means
there is no entanglement at $T=0$. However by increasing $T$, the
entangled state $|\phi_{3}\rangle$ and $|\phi_{4}\rangle$ will mix
with the state, which makes the entanglement increase. From the
two figures in Fig.5, we can see that when $b$ is raised, the
critical magnetic field $B_{c}$ increases, but the maximum
entanglement value that the system can arrive becomes smaller.

\section{IV. The conclusions } In conclusion, we have investigate the thermal
entanglement in two qubit Heisenberg \emph{XXZ} spin chain under
an inhomogeneous magnetic field. The ground-state entanglement and
thermal entanglement at a finite temperature are given. We find
the entanglement exists for both antiferromagnetic and
ferromagnetic cases. And the entanglement is enhanced by
increasing the interaction of \emph{z}-component $J_{z}$. The
critical inhomogeneous magnetic field is independent of $J_{z}$.
The critical magnetic field $B_{c}$ increases with the increasing
$|b|$ but the maximum entanglement value that the system can
arrive becomes smaller.

\end{document}